\documentclass[aps,prl,reprint,amsmath,amssymb,superscriptaddress]{revtex4-2}

\usepackage{graphicx}
\usepackage{dcolumn}
\usepackage{bm}
\usepackage{xcolor}
\usepackage{braket}

\begin{document}


\title{A Universal Four-Fermion Formation Framework and Odd-Even Staggering in $\alpha$ Decay}


\author{Boshuai Cai}
\email[]{caibsh@mail.sysu.edu.cn}
\affiliation{Sino-French Institute of Nuclear Engineering and Technology, Sun Yat-sen University, Zhuhai, 519082, China}

\author{Cenxi Yuan}
\email[]{yuancx@mail.sysu.edu.cn}
\affiliation{Sino-French Institute of Nuclear Engineering and Technology, Sun Yat-sen University, Zhuhai, 519082, China}

\author{Chong Qi}
\email[]{chongq@kth.se}
\affiliation{Department of Physics, Royal Institute of Technology, Stockholm, SE-10691, Sweden}
\affiliation{Sino-French Institute of Nuclear Engineering and Technology, Sun Yat-sen University, Zhuhai, 519082, China}


\date{\today}

\begin{abstract}
Clustering phenomena are common in many physical systems across multiple scales. The nuclear $\alpha$ decay is one of the earliest observed evidences of clustering in quantum systems, yet its formation process remains poorly understood even today. In this letter, we propose a novel global odd-even staggering (OES) feature in $\alpha$ decay, which emerges during the clustering process. To unveil its origin, we develop a Universal Four-Fermion Formation Framework (U4F), which describes the formation of any four-nucleon cluster, such as $\alpha$ particle, from a general microscopic wave function, without assuming the preexistence of clustering or pairing. By combining U4F with the large-scale configuration-interaction approach, we demonstrate that the OES effect in $\alpha$ decay arises from the suppression of clustering correlations due to unpaired nucleons. These findings significantly advance our understanding of cluster formation in nuclei and have important implications for the production of new elements and nuclear synthesis in the universe.

\end{abstract}


\maketitle


\textit{Introduction}\textemdash Clustering is a significant phenomenon observed across various physical systems, from the macroscopic scales of galaxies \cite{springel_simulations_2005} to the microscopic interactions between nucleons \cite{ebran_how_2012}. In nuclei, $\alpha$ decay is the earliest observed \cite{rutherford_viii_1899, rutherford_emanations_1901} and the most prominent manifestation of clustering within a quantum many-body system. The process of $\alpha$ decay is typically described in two steps: the formation of the $\alpha$ particle within the nucleus and its subsequent emission through the Coulomb and centrifugal barriers via a quantum tunneling process \cite{lane_r-matrix_1958, qi_alpha_2016, auranen_superallowed_2018}. While the latter is well understood, the mechanism underlying the formation of the $\alpha$ particle remains elusive \cite{qi_recent_2019, ren_alpha-clustering_2018, delion_alpha-clustering_2023, zhao_microscopic_2023}. Although nucleon-nucleon correlations and configuration mixing are widely regarded as key ingredients \cite{harada_alpha-particle_1961, mang_shell_1962, rasmussen_simplified_1963, poggenburg_theoretical_1969, ichimura_alpha-particle_1973, tonozuka_surface_1979, janouch_role_1982, dodig-crnkovic_exact_1989, insolia_microscopic_1991,  delion_anisotropy_1992, lovas_microscopic_1998, patial_microscopic_2016, varga_absolute_1992, id_betan_alpha_2012, dong_alpha-cluster_2021}, the complexity of incorporating full correlations often restricts the analysis to a subspace formed by the direct product of proton-pair and neutron-pair states with specific angular momentum symmetry, which inevitably omits crucial correlations \cite{ tonozuka_surface_1979, lovas_microscopic_1998, qi_recent_2019}.

The pairing correlation plays a crucial role in nearly all nuclear phenomena \cite{dean_pairing_2003, zelevinsky_nuclear_2003, tanihata_models_2022, cai_random_2024}, contributing to the systematic odd-even staggering (OES) observed in nuclear properties such as masses \cite{satula_odd-even_1998, mougeot_mass_2021, changizi_empirical_2015} and charge radii \cite{marsh_characterization_2018, de_groote_measurement_2020, koszorus_charge_2021}. Since the formation of the $\alpha$ particle is closely related to pairing correlations \cite{soloviev_effect_1962, bohr1998nuclear, andreyev_signatures_2013, seif_nucleon_2015}, it is expected that $\alpha$ decay also exhibits generic OES. Recent studies have shown the OES effect in the decay energy ($Q_\alpha$), half-life ($T_\alpha$), and $\alpha$ formation probability ($F_\alpha$) in certain isotopic chains \cite{yang_new_2022, xu_favored_2005, yang_examining_2022, delion_even-odd_2016, sun_systematic_2017, deng_correlation_2021}, while the existence of a global OES relationship remains unclear.

This raises two critical questions regarding $\alpha$ decay: 1) Is OES a global feature of $\alpha$ formation across the nuclear landscape? 2) How can we describe the $\alpha$ formation including full correlations? In this letter, we argue that the novel global OES feature arises solely from the $\alpha$ formation process. To further understand this feature, we construct a novel universal four-fermion formation framework (U4F), which does not rely on preexisting clustering or specific symmetry assumptions. Using Po and At isotopes as examples, we demonstrate how the U4F evaluates both the $\alpha$ formation and quartet energy, which accounts for the observed OES effect.

\textit{Global OES feature}\textemdash To address the first question, we begin by analyzing the systematic OES properties of $Q_\alpha$, $T_\alpha$, and $F_\alpha$. The first two are the most direct observables in $\alpha$ decay, while the latter provides a scaled measure of the $\alpha$ formation probability. Each of these three quantities exhibits local OES feature in $\alpha$ decay \cite{yang_new_2022, xu_favored_2005, yang_examining_2022, delion_even-odd_2016, sun_systematic_2017, deng_correlation_2021}. A comparison among them helps identify a global OES feature and its potential origin, which may stem from the formation, tunneling, or overall decay process. The three-point indicator $\Delta$ (see, for example, \cite{yang_new_2022, de_groote_measurement_2020}) is widely used to quantify OES in isotopic and isotonic chains. If a physical quantity $X$ exhibits ideal OES along an isotopic chain, its three-point indicator $\Delta_{(n)}X = \frac{2X(Z,N) - X(Z,N-1) - X(Z,N+1)}{2}$ for odd-$N$ and even-$N$ isotopes should display opposite signs. A similar pattern is expected along isotonic chains.

\begin{figure}[htp]
 \begin{center}
\includegraphics[width=0.49\textwidth]{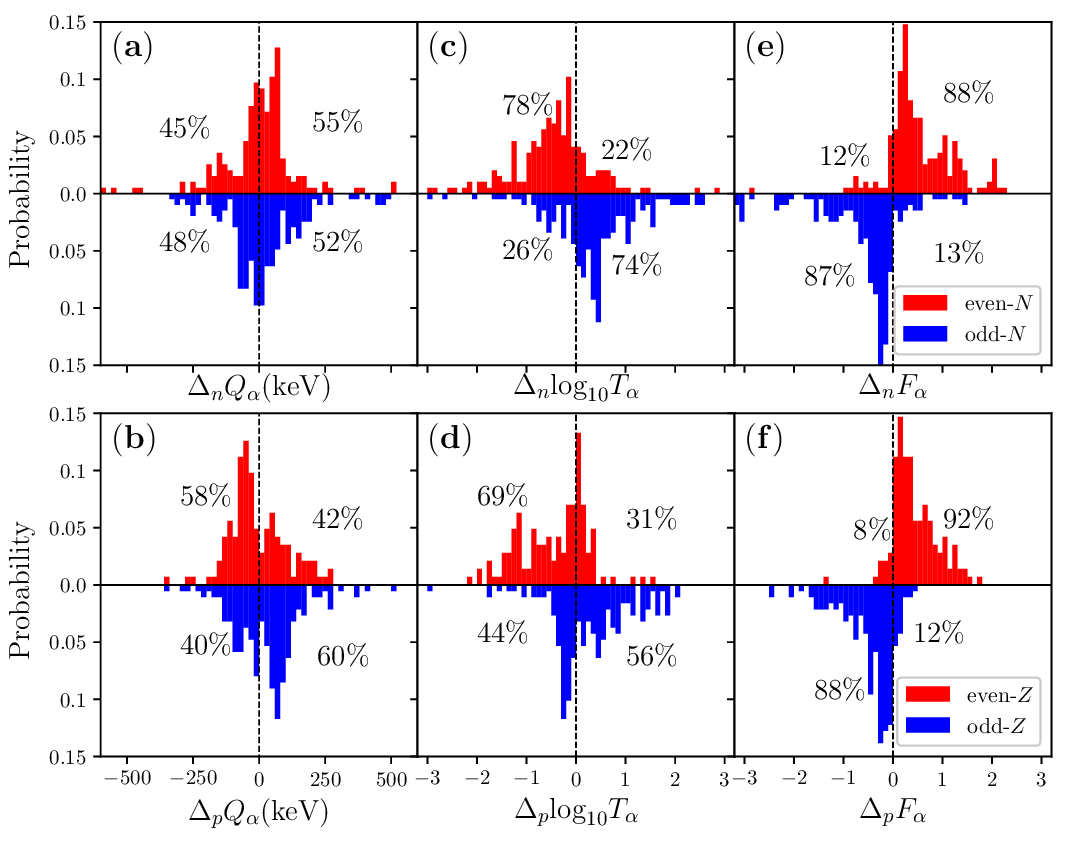}
\end{center}
\caption{\label{fig:delta}Distribution of $\Delta Q_\alpha$, $\Delta \log_{10}T_\alpha$, and $\Delta F_\alpha$ for nuclides with even-$N$, odd-$N$, even-$Z$, and odd-$Z$. The upper (lower) panels display data for isotopic (isotonic) chains, denoted as $\Delta_{n(p)}$. Each panel also indicates the proportions of positive and negative values of $\Delta$. The experimental values of $Q_\alpha$ and $T_\alpha$ are collected from AME2020 \cite{wang2021ame2020} and NUBASE2020 \cite{kondev_nubase2020_2021}, respectively.}
\end{figure}

As shown in Figs. \ref{fig:delta}(a-b), the distribution of $\Delta_{n (p)} Q_\alpha$ for nuclides with even and odd $N (Z)$ values centers around zero. Although $Q_\alpha$ shows strong staggering in the northwest region of $^{208}$Pb \cite{yang_new_2022}, no global OES characteristic is found for $Q_\alpha$. In the case of $T_\alpha$,  Fig. \ref{fig:delta}(c) presents that the blocking of unpaired neutrons results in a more widespread presence of OES along isotopic chains compared with $Q_\alpha$, which is consistent with previous work \cite{xu_favored_2005, yang_examining_2022}. When this trend is less pronounced along isotonic chains in Fig. \ref{fig:delta}(d), particularly quasi-half of $\Delta_{p}\log_{10}T_\alpha$ of odd-$Z$ nuclide are negative, it is far from confirming the blocking of unpaired protons on $T_\alpha$. 
It is not surprising that $T_\alpha$ shows a more pronounced OES feature than $Q_\alpha$, as the decay process involves both formation and tunneling processes.

\begin{figure}[htp]
\includegraphics[width=0.49\textwidth]{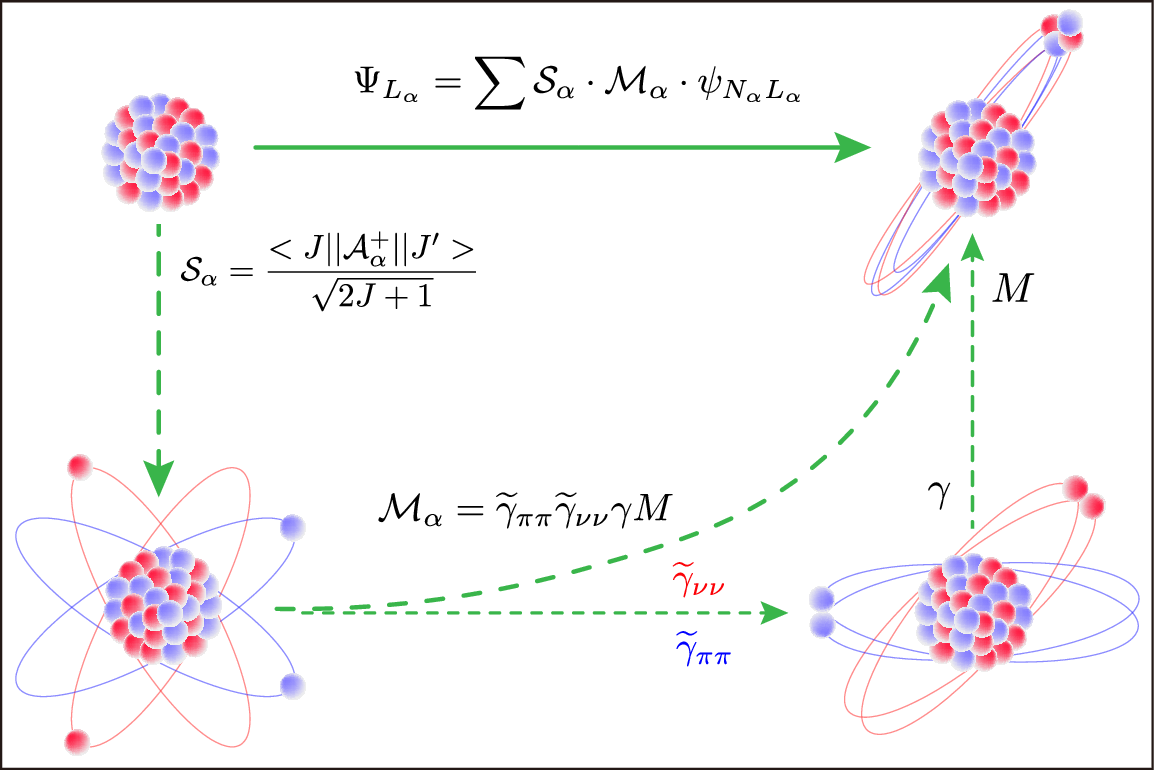}
\caption{\label{fig:mechanism} Illustration of the microscopic mechanism of the formation process of an $\alpha$ particle in motion of the $L_\alpha$ wave in nucleus. Key components include the selection of orbitals, represented by the quartet amplitude $\mathcal{S}_\alpha$, and the center of mass mapping process. The latter includes transformations such as the $jj$ to $ls$ transformation $\gamma$ and $\widetilde{\gamma}$, as well as the Talmi-Moshinsky transformation $M$.}
\end{figure}

As a generalization of the Geiger-Nuttall law, the universal decay law (UDL), $\log_{10}T_{1/2} = a\chi' + b\rho' + c$, where $a$, $b$ and $c$ are constants, has been presented to explain the general relation between the half-life and the decay energy of charged cluster emission \cite{qi_universal_2009, qi_microscopic_2009}. Its ability of describing $\alpha$ decay has been validated through systematic analysis \cite{cai_2020}. When the decay energy appears only in the term $a\chi'=aZ_\alpha Z_d\sqrt{A_\alpha A_d/(A_\alpha+A_d)Q_\alpha}$, which is contributed by quantum tunneling, the rest terms $b\rho'+c = b\sqrt{Z_\alpha Z_d(A_\alpha^{1/3}+A_d^{1/3})A_\alpha A_d/(A_\alpha+A_d)} +c$ are related with the cluster formation, where $A_\alpha$ and $A_d$ ($Z_\alpha$ and $Z_d$) are the mass (charge) numbers of the $\alpha$ particle and daughter nucleus, respectively. Thus, the expression $a\chi' - \log_{10}T_\alpha$ effectively decouples the tunneling contribution from $T_\alpha$, thereby providing an estimate of $F_\alpha$. 
 As shown in Fig. \ref{fig:delta}(e-f), about 90\% of nuclides with even $N$ and $Z$ exhibit a positive $\Delta_{n(p)} F_\alpha$, whereas the majority of odd-$N$ $(Z)$ emitters display negative values. This dramatic distinction, compared with the systematic trends in $Q_\alpha$ and $T_\alpha$, suggests the existence of a global OES feature in $\alpha$ particle formation.

While the extraction of $\Delta F_\alpha$ above is linked to $T_\alpha$ and $Q_\alpha$, and most previous microscopic studies systematically underestimated the $\alpha$ formation amplitude, it has remained unclear whether the global OES feature arises from the formation process itself or is merely a residual effect of quantum tunneling.
By developing a novel universal four-fermion formation framework (U4F), we propose this global OES feature in $\alpha$ decay originating from the formation process itself, which generalizes and clarifies previous understandings.


\textit{U4F}\textemdash Describing $\alpha$ clustering from a microscopic perspective has long posed a significant challenge \cite{qi_recent_2019, ren_alpha-clustering_2018, delion_alpha-clustering_2023, zhao_microscopic_2023}. The configuration-interaction approach is typically used to obtain the wave functions of both the initial and final states, as it accounts for multi-fermion correlations beyond the mean field \cite{lovas_microscopic_1998}. Many efforts have been made since the 1960s to deduce $\alpha$ clustering formation using configuration-interaction approaches \cite{harada_alpha-particle_1961, mang_shell_1962, rasmussen_simplified_1963, poggenburg_theoretical_1969, ichimura_alpha-particle_1973, tonozuka_surface_1979, janouch_role_1982, dodig-crnkovic_exact_1989, insolia_microscopic_1991,  delion_anisotropy_1992, lovas_microscopic_1998, patial_microscopic_2016, varga_absolute_1992, id_betan_alpha_2012, dong_alpha-cluster_2021}, but largely limited by constraints in symmetry. For instance, many works consider only $J=0$ $pp$ and $nn$ pairs. The quartet configurations, which involve the full coupling of four fermions, have not been adequately extracted from the configuration-interaction wave function, while triplet configurations have been successfully obtained \cite{navratil_cluster_2004}.

In essence, the challenge lies in extracting multi-fermion correlations from the complex nuclear wave function, which is usually expressed in an independent-particle basis, and mapping them onto a localized four-fermion cluster defined in its intrinsic frame. In this letter, we confront this challenge by developing U4F that directly evaluates four-fermion clustering from the microscopic wave function in two steps. In the first step, we identify the two- and four-body correlations that contribute to the formation of four-fermion cluster in the frame of the parent nucleus, thereby and quantify the relevant quartet configurations and their amplitudes. In the second step, we perform a transformation from the parent nucleus frame to the intrinsic frame of the quartet, mapping the quartet onto the escaping cluster. In principle, the quartet amplitude can be extracted from shell-model wave functions, with the mapping carried out via the Talmi-Moshinsky transformation. Most existing studies have been restricted to $\alpha$-emitters near shell closures, including in particular $^{212}$Po. The consistent integration of quantum harmonic oscillator wave function, Talmi-Moshinsky transformation and quartet amplitude into a large-scale, parameter-free framework for alpha formation has not been realized before due to its theoretical and numerical difficulties.

In practice, the identification of the quartet configurations is studied by defining a quartet amplitude as
\begin{equation}
\begin{aligned}
\mathcal{S}_\alpha({J\{|^{k_1k_2(j_{12})}_{k_3k_4(j_{34})}(J_\alpha);J'}) = \frac{\langle J||\mathcal{A}^{(J_\alpha)+}_{k_1k_2(j_{12})k_3k_4(j_{34})}|| J'\rangle}{\sqrt{2J+1}},
\end{aligned}
\end{equation}
where $\mathcal{A}^{(J_\alpha)+}_{k_1k_2(j_{12})k_3k_4(j_{34})} = [[a_{k_1}^+\otimes a_{k_2}^+]^{(j_{12})} \otimes[a_{k_3}^+\otimes a_{k_4}^+]^{(j_{34})} ]^{(J_\alpha)}$ 
is the irreducible four-nucleon creation operator, and the subscript $\alpha$ indicates the quartet. For simplicity, we use $k_i$ to represent the radial quantum number $n_i$, isospin $t_i$, spin $s_i$, orbital $l_i$, and total angular momentum $j_i$. $|\mathcal{S}_\alpha|^2$ quantifies the probability that two fermions in orbitals $k_1$ and $k_2$ (coupled to $j_{12}$), and two fermions in orbitals $k_3$ and $k_4$ (coupled to $j_{34}$) contribute to the formation of a specific quartet which eventually forms an four-fermion cluster with $J_\alpha$ in the parent nucleus. Here, the initial (final) state of the parent (daughter) nucleus has spin $J$ ($J'$).  Clustering is the correlated contribution from all possible quartet amplitudes.

For the second step, the dynamic evolution of the four fermions occupying orbitals $k_{i, i \in [1,4]}$ to form a cluster is related to the transformation from the laboratory frame to the intrinsic frame of the four-fermion cluster. To achieve this, we first transform the initial single-particle wave functions of the four fermions from the $jj$ coupling to the $ls$ coupling, which isolates the effects of the radial wave function and intrinsic spin. Then, we transform the radial wave functions in $ll$ coupling to the intrinsic frame of the quartet through the Talmi-Moshinsky transformation and separate out the center of mass motion of the quartet. After these transformations, we obtain the wave function of the $\alpha$ particle in each quartet configuration
\begin{equation}
\begin{aligned}
\varPhi_{J_\alpha M_\alpha}^{k_1k_2(j_{12})k_3k_4(j_{34})}
= \sum_{ l_{12} s_{12}  l_{34} s_{34}} \sum_{N_\alpha L_\alpha n_\alpha l_\alpha} \gamma^{(J_\alpha)}_{J_\alpha 0}(^{j_{12} l_{12} s_{12}}_{j_{34} l_{34} s_{34}}) \cdot \\
 \widetilde\gamma^{(j_{12}t_{12})}_{l_{12} s_{12}}(^{j_1 l_1 s_1}_{j_2 l_2 s_2}) \cdot \widetilde\gamma^{(j_{34}t_{34})}_{l_{34} s_{34} }(^{j_3 l_3 s_3}_{j_4 l_4 s_4})   \cdot 
M_l(^{N_\alpha L_\alpha n_\alpha l_\alpha}_{n_{12}l_{12} n_{34}l_{34}}) \cdot \\
[ [ [\psi_{N_\alpha L_\alpha} \otimes \psi_{n_\alpha l_\alpha}]_{nl} \otimes [\psi_{0s} \otimes \psi_{0s}]_{0s}  ]_{nl} \otimes \\
[ [\chi_{s_1} \otimes \chi_{s_2}]_{s_{12}} \otimes [\chi_{s_3} \otimes \chi_{s_4}]_{s_{34}}  ]_{S} ]^{J_\alpha}_{M_\alpha},
\end{aligned}
\end{equation}
where $\psi$ denotes the single nucleon wave function, $\chi$ denote the spin wave function, 
$l_{12 (34)} = l_{1(3)} \otimes l_{2 (4)}$ is the total orbital angular momentum of the nucleon-pair, 
$s_{12 (34)} = s_{1 (3)} \otimes s_{2 (4)}$ is the total intrinsic spin of the nucleon-pair, 
$t_{12 (34)} = t_{1 (3)} \otimes t_{2 (4)}$ is the total isospin of the nucleon-pair, 
$S = s_{12}\otimes s_{34}$ is the total intrinsic spin of the quartet, 
$l = l_{12}\otimes l_{34}$ is the total orbital angular momentum of the quartet, 
$L_\alpha$ is the orbital angular momentum of the center of mass of the quartet,
$l_\alpha$ is the orbital angular momentum of the relative motion of the two nucleon-pairs,
$2N_\alpha + L_\alpha + 2n_\alpha + l_\alpha = 2n_{12} + l_{12} + 2n_{34} + l_{34} = 2(n_1+n_2+n_3+n_4) + l_1+l_2+l_3+l_4$,
$\gamma$ is the $jj$ to $ls$ transformation coefficient \cite{talmi_simple_1993}, $\widetilde{\gamma}$ is the exchange asymmetric $jj$ to $ls$ transformation coefficient, 
$\widetilde\gamma^{(j_{12}t_{12})}_{l_{12} s_{12}}(^{j_1 l_1 s_1}_{j_2 l_2 s_2}) = \frac{1- (-1)^{t_{12}+s_{12}}}{\sqrt{2(1+\delta_{k_1k_2})}} \gamma^{(j_{12})}_{l_{12} s_{12}}(^{j_1 l_1 s_1}_{j_2 l_2 s_2})$.
$M$ is the Brody-Moshinsky Bracket, as used in the Talmi-Moshinsky transformation \cite{talmi_simple_1993}.
The multiplication of these transformation coefficients in each coupling channel can be noted as $\mathcal{M}_\alpha(^{N_\alpha L_\alpha} _{n_\alpha l_\alpha} \{|^{k_1k_2(j_{12}l_{12}s_{12})}_{k_3k_4(j_{34}l_{34}s_{34})} )$.

Given the possible configurations, the coefficient of the wave $\psi_{N_\alpha L_\alpha}$ of the $\alpha$ particle formed by two protons occupying orbitals $k_1$ and $k_2$ (coupled to $j_{12}$) and two neutrons occupying orbitals $k_3$ and $k_4$ (coupled to $j_{34}$) is,
\begin{equation}
\mathcal{S}_\alpha({J\{|^{k_1k_2(j_{12})}_{k_3k_4(j_{34})}(J_\alpha);J'}) \cdot \mathcal{M}_\alpha(^{N_\alpha L_\alpha} _{n_\alpha l_\alpha} \{|^{k_1k_2(j_{12}l_{12}s_{12})}_{k_3k_4(j_{34}l_{34}s_{34})}).
\end{equation}
The entire process is illustrated in Fig. \ref{fig:mechanism}. When the $\alpha$ particle emerges on the surface of the parent nucleus, the nuclear force becomes negligible. The wave function is dominated by the Coulomb field and must match a Coulomb outgoing wave with angular momentum $L_\alpha$ (i.e., the Coulomb Hankel function $H^+_{L_\alpha}$). This condition provides the theoretical formation probability of the $\alpha$ particle 
\begin{equation}
\begin{aligned}
\label{f_th}
|\mathcal{F}_{L_\alpha}|^2_{\text{th.}} 
= \sum_{k_{i,i\in[1,4]}, j_{12}, j_{34}} \sum_{k_{i,i\in[1,4]}', j_{12}', j_{34}'}  \sum_{l_{12}s_{12}l_{34}s_{34}} \sum_{l_{12}'s_{12}'l_{34}'s_{34}'} \\
\sum_{N_\alpha n_\alpha l_\alpha}\mathcal{S}_\alpha({J\{|^{k_1k_2(j_{12})}_{k_3k_4(j_{34})}(J_\alpha);J'}) \cdot  \mathcal{S}_\alpha({J\{|^{k_1'k_2'(j_{12}')}_{k_3'k_4'(j_{34}')}(J_\alpha);J'}) \cdot \\
\mathcal{M}_\alpha(^{N_\alpha L_\alpha} _{n_\alpha l_\alpha} \{|^{k_1k_2(j_{12}l_{12}s_{12})}_{k_3k_4(j_{34}l_{34}s_{34})} \cdot  \mathcal{M}_\alpha(^{N_\alpha L_\alpha} _{n_\alpha l_\alpha} \{|^{k_1'k_2'(j_{12}'l_{12}'s_{12}')}_{k_3'k_4'(j_{34}'l_{34}'s_{34}')} \cdot\\
\psi_{N_\alpha L_\alpha}^2(R_\alpha)
\end{aligned}
\end{equation}
and the experimental formation probability
\begin{equation}
\label{f_exp}
|\mathcal{F}_{L_\alpha}|^2_{\text{expt.}} = \frac{\mu}{\hbar R} \frac{|H^+_{L_\alpha}(kR)|^2}{kR} \frac{1}{T_\alpha}
\end{equation}
where $\mu$ represents the reduced mass of the $\alpha$ particle and the daughter nucleus, $k=\sqrt{2\mu Q_\alpha}/\hbar$ is the wavenumber, $R=r_0(A_\alpha^{1/3}+A_d^{1/3})$ denotes the distance between the $\alpha$ particle and the daughter nucleus beyond the influence of the nuclear potential \cite{qi_universal_2009, delion_systematics_2006}, $R_\alpha$ is the radial coordinate of the $\alpha$ particle, and $R_\alpha = A_dR/(A_\alpha+A_d)$. The radial wave function $\psi_{N_\alpha L_\alpha}$ of the $\alpha$ particle is expressed using the form of the quantum harmonic oscillator whose size parameter is four times that of a single nucleon, as indicated by the harmonic oscillator property \cite{talmi_simple_1993}.

 \begin{figure}[htpb]
 \begin{center}
 \includegraphics[width=0.48\textwidth]{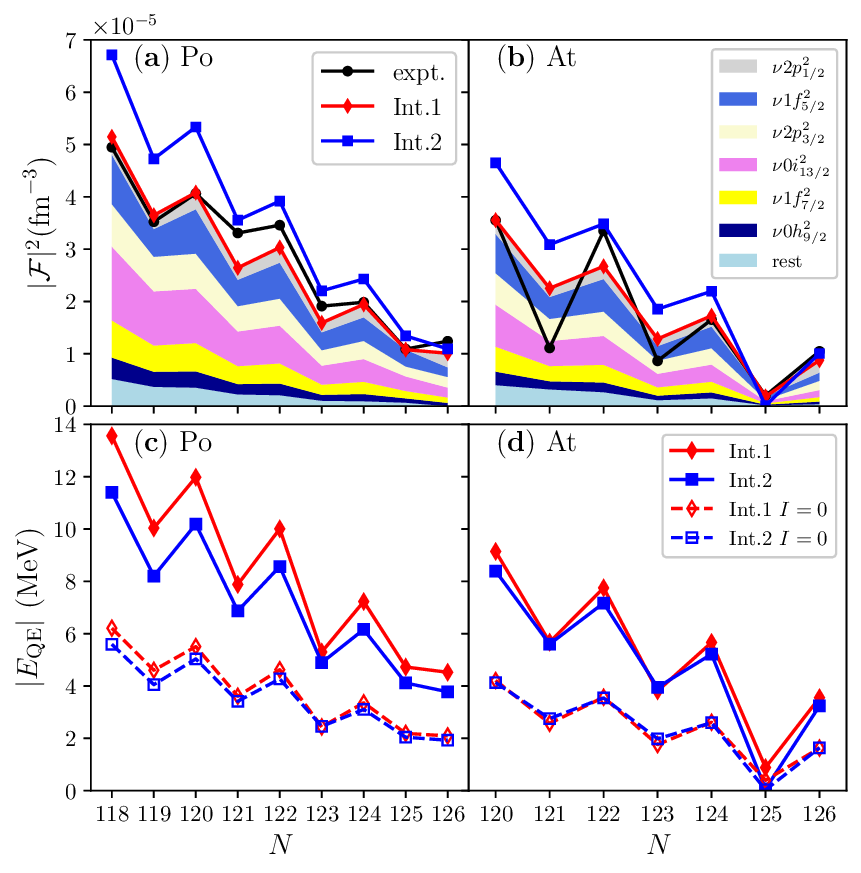}
\end{center}
\caption{\label{fig:contrast_p} Panels ($a$-$b$): The $\alpha$ formation probability of isotopes $^{202-210}$Po and $^{205-211}$At, both extracted from experimental data by Eq. (\ref{f_exp}) and estimated theoretically by Eq. (\ref{f_th}) using two groups of Hamiltonian (Int.1 and Int.2). Panels ($c$-$d$): Total Quartet Energy and the pairing correlation energy from like-particle spin-zero  ($I=0$) pairs of Po and At isotopes extracted according to Eq. (\ref{eq:QE}).}
\end{figure}

\textit{Large-scale Description}\textemdash 
Within the framework developed above, we study the formation probability of $\alpha$ particle of Po and At isotopes as examples. Po isotopes have two protons beyond the magic number $Z=82$ and are normally taken as the typical $\alpha$ decay, while At isotopes has one proton more than Po isotope to show the OES in isotopes with odd number of proton. We focus on the $\alpha$ decay channel from ground states of $^{202-210}$Po and $^{205-211}$At to final states of their daughter nuclei with identical spin and parity, resulting in the $s$-wave motion of the emitted $\alpha$ particle with $L_\alpha=0$. Large-scale configuration-interaction shell model are performed within the $h_9fpi_{13}$ model space, which includes six orbitals for both protons and neutrons: $0h_{9/2}$, $1f_{7/2}$, $0i_{13/2}$, $2p_{3/2}$, $1f_{5/2}$, and $2p_{1/2}$. 
Two groups of Hamiltonian are applied. One is derived from the modified Kuo-Herling interaction \cite{warburton_first-forbidden_1991, warburton_appraisal_1991}, the monopole-based universal interaction (VMU) \cite{otsuka_novel_2010} and the M3Y spin-orbit interaction \cite{bertsch1977}, as referenced in previous studies \cite{zhang_new_2021, yang_new_2022}.
Another is constructed only by VMU and M3Y spin-orbit interaction. These two groups of Hamiltonian are noted as Int.1 and Int.2, respectively.
These studies, performed through the KSHELL code \cite{kshell2019}, are essential for determining the binding energies and wave functions of the initial and final states in the $M$-scheme, which are necessary for investigating the quartet amplitude $\mathcal{S}_\alpha$.

The theoretical $\alpha$-formation probabilities, calculated from large-scale configuration-interaction wave functions using Eq.~(\ref{f_th}), are shown in Fig.~\ref{fig:contrast_p}(a–b) and compared with values extracted from experimental half-lives via Eq.~(\ref{f_exp}). Overall, the results agree well with experiment and reproduce two essential features. 
First, the $\alpha$-formation probability decreases toward the $N=126$ shell closure, reflecting the reduction of pairing correlations near the closed shell \cite{andreyev_signatures_2013, qi_validity_2014}. 
Second, the calculations accurately capture the OES feature in $\alpha$ formation. The $\alpha$ particle can be viewed as formed by correlated neutron and proton Cooper pairs, induced by strong pairing correlations. In odd-$N$ nuclei, the unpaired neutron reduces not only one collective pairs but also the total pairing correlation, as the blocking of the odd particle decreases the phase space for the formation of pairs.
Furthermore, an analysis of neutron-pair configurations, illustrated with Int.1, shows that $\alpha$ formation in Po and At isotopes is dominated by configurations where the two neutrons occupy the same orbital. The OES originates primarily from these configurations, as the blocked neutron in odd-$N$ nuclei reduces the available pairing correlations within each single-$j$ orbital.

The nuclear binding energy reflects the multi-nucleon correlation within a nucleus, where effective neutron and proton pairing gaps can be extracted via various mass filters like the three-point formula $\frac{(-1)^N(2\mathrm{BE}(Z, N) - \mathrm{BE}(Z, N+1) - \mathrm{BE}(Z, N-1))}{2}$ \cite{broglia_fifty_2013,duguet2001,changizi2015}, which estimates the difficulty of breaking a neutron-pair (equivalent to the strength of neutron-pair formation) within the nucleus. 
We have extracted the pairing gap using the three-point formula from experimental binding energies  \cite{wang2021ame2020} and those calculated by the configuration-interaction shell model. A stricking feature is that they indeed exhibit the same shell effect and OES features as the $\alpha$ formation probability. This similarity suggests that the OES observed in the $\alpha$ formation probability of these Po and At isotopes is closely tied to the pairing correlation.
However, the  pairing gaps thus extracted from the experimental binding energy can be contaminated by single-particle or mean field contributions \cite{duguet2001,changizi2015}. 

The U4F provides an approach to probe the correlation between $\alpha$ formation and interaction through the quartet energy
\begin{equation}
\label{eq:QE}
E_{\mathrm{QE}}= \bra{J_\alpha} V \ket{J_\alpha},
\end{equation}
where $V$ denotes the nucleon-nucleon interaction and $\ket{J_\alpha} = \sum_{k_{i, i\in[1,4]}}^{j_{12} j_{34}}  \mathcal{S}_\alpha (J\{| ^{k_1k_2(j_{12})}_{k_3k_4(j_{34})}(J_\alpha);J' ) \cdot  \ket{J_\alpha \{| ^{k_1k_2(j_{12})}_{ k_3k_4(j_{34})}}$ represents the quartet wave function. The magnitude of $E_{\mathrm{QE}}$ quantifies the strength of interaction-driven quartet correlations, with larger $|E_{\mathrm{QE}}|$ corresponding to stronger quartet correlation. To disentangle different contributions, we separate the pairing channel, where the extracted pairing energy directly reflects genuine pairing correlations, free from the mean-field contamination inherent in traditional mass-filter analyses. 
As illustrated in FIG. \ref{fig:contrast_p}(c-d), both the total $|E_{\mathrm{QE}}|$ and the total pairing energy contribution exhibit the same shell feature and OES feature as the $\alpha$ formation probability. This also demonstrates that the U4F proposed here offers a robust framework to link $\alpha$ formation with nucleon–nucleon interactions.


\textit{Conclusion}\textemdash The OES in $\alpha$ formation probability is identified as a global feature across the nuclear landscape. We propose the U4F, a microscopic framework for four-fermion formation from complex nuclear many-body wave functions that can account for full symmetry and nucleon correlations. As an example, this framework is validated both qualitatively and quantitatively through the OES and evolution of $\alpha$ formation probability in Po and At isotopes. The core principles of this framework can be extended to explore heavier clustering within the nucleus, potentially advancing our understanding of multi-nucleon transfer and cluster capture reactions. Since $\alpha$ decay plays a pivotal role in identifying new nuclides, particularly in the heavy region of the nuclear landscape, this framework provides valuable insights into the formation of new isotopes \cite{zhang_new_2021,yang_discovery_2024} and elements. Moreover, given that $\alpha$ decay and $\alpha$ capture are inverse processes, this framework may significantly enhance our understanding of $\alpha$-capture reaction in stellar nucleosynthesis \cite{elhatisari_ab_2015, shen_alpha-cluster_2021}.

\begin{acknowledgments}
This work has been supported by the Guangdong Major Project of Basic and Applied Basic Research under Grant No. 2021B0301030006, the National Natural Science Foundation of China under Grant No. 12475129, the computational resources from Sun Yat-sen University, and the National Supercomputer Center in Guangzhou.
\end{acknowledgments}

\bibliography{alphamapping-bibliography}

\begin{widetext}
\appendix*
\section{Supplemental Material}

The supplemental material includes: 1) the quartet amplitude, 2) the center of mass amplitude, and 3) the decay process.

\subsection{I. Quartet Amplitude}\label{sec:mt:ata}

The quartet amplitude ($\mathcal{S}_\alpha$) is defined through the two-protons-two-neutrons creation operator linking the final $(A-4)$-nucleons state $|\omega' J'\rangle$ to the initial $A$-nucleons state $|\omega J\rangle$, where $J$ denotes spin and $\omega$ denotes other quantum number. In $j$-scheme, it is expressed by 
\begin{equation}
\mathcal{S}_\alpha({J\{|^{k_1k_2(j_{12})}_{k_3k_4(j_{34})}(J_\alpha);J'}) = 
\frac{\langle\omega J||\mathcal{A}^{(J_\alpha)+}_{k_1k_2(j_{12})k_3k_4(j_{34})}||\omega' J'\rangle}{\sqrt{2J+1}},
\end{equation}
where $k_{i, i\in[1,4]}$ means the single particle orbital $n_il_ij_i$ occupied by the two protons and two neutrons; $J_\alpha = j_{12}\otimes j_{34} = [j_1\otimes j_2]\otimes [j_3\otimes j_4]$ is the spin coupled by the four nucleons, and constrained also by the initial and final states, i.e., $J = J_\alpha \otimes J'$; $\mathcal{A}^{(J_\alpha)+}_{k_1k_2(j_{12})k_3k_4(j_{34})}$ is the tensor operator creating two protons and two neutrons in orbitals $k_{i, i\in[1,4]}$ coupling to $J_\alpha$. $|\mathcal{S}_\alpha|^2$ thus represents the probability that two protons in orbitals $n_1l_1j_1$ and $n_2l_2j_2$ coupling to $j_{12}$ and two neutrons in orbitals $n_3l_3j_3$ and $n_4l_4j_4$ coupling to $j_{34}$ in the parent nucleus participate the formation of an $\alpha$ particle with spin $J_\alpha$. 

Applying the Wigner-Eckart theorem \cite{wigner_gruppentheorie_1931, eckart_application_1930}, the definition of $\mathcal{S}_\alpha$ in $m$-scheme is deduced to be 
\begin{equation}
\begin{aligned}
\mathcal{S}_\alpha({J\{|^{k_1k_2(j_{12})}_{k_3k_4(j_{34})}(J_\alpha);J'}) &= \frac{\langle\omega J||\mathcal{A}^{(J_\alpha)+}_{k_1k_2(j_{12})k_3k_4(j_{34})}||\omega' J'\rangle}{\sqrt{2J+1}} \\
&= \frac{(-1)^{(J-M)}\langle\omega JM|\mathcal{A}^{(J_\alpha M_\alpha)+}_{k_1k_2(j_{12})k_3k_4(j_{34})}|\omega'J'M'\rangle}{\begin{pmatrix}
J & J_\alpha & J'\\
-M & M_\alpha & M' 
\end{pmatrix}},
\end{aligned}
\end{equation}
where $\mathcal{A}^{(J_\alpha M_\alpha)+}_{k_1k_2(j_{12})k_3k_4(j_{34})}=[[a_{k_1}^+ \otimes a_{k_2}^+]^{(j_{12})} \otimes [a_{k_3}^+ \otimes a_{k_4}^+]^{(j_{34})}]_{M_\alpha}^{J_\alpha} $ is the component of tensor operator $\mathcal{A}^{(J_\alpha)+}_{k_1k_2(j_{12})k_3k_4(j_{34})}$, with $M_\alpha=M-M'$ and $a^+_{k_i, i\in[1,4]}$ the operator creating a nucleon in orbital $k_i$. In this way, $\mathcal{S}_\alpha$ includes two parts: one is determined by the structure (configuration) of the initial and final states, and another is intrinsically determined by the coupling coefficient.

\subsection{II. Center of Mass Amplitude}\label{sec:mt:cmma}

The center of mass amplitude ($\mathcal{M}_\alpha$) is the amplitude that two protons and two neutrons in orbitals $(n_il_ij_is_it_i)_{i\in[1,4]}$ form an $\alpha$ particle of which the center of mass is in orbital $N_\alpha L_\alpha$, the relative motion of proton-proton pair and neutron-neutron pair is in orbital $n_\alpha l_\alpha$, where $s_i$ ($t_i$) represents spin (isospin). To deduce $\mathcal{M}_\alpha$, one has to consider processes of:  
\begin{itemize}
\item[1)] Two nucleons in orbitals $n_1l_1j_1s_1t_1$ and $n_2l_2j_2s_2t_2$ form a di-nucleon in orbital $n_{12}l_{12}j_{12}s_{12}t_{12}$, where $n_{12}$ is the radial quantum number, $l_{12}$ is the orbital angular momentum, $j_{12}$ is the total spin, $s_{12}$ is the intrinsic spin, and $t_{12}$ is the isospin of the di-nucleon; 
\item[2)] Two nucleons in orbitals $n_3l_3j_3s_3t_3$ and $n_4l_4j_4s_4t_4$ form another di-nucleon in orbital $n_{34}l_{34}j_{34}s_{34}t_{34}$;
\item[3)] Two di-nucleons in orbitals $n_{12}l_{12}j_{12}s_{12}t_{12}$  and $n_{34}l_{34}j_{34}s_{34}t_{34}$ form a quartet in orbital $nl J_\alpha ST$;
\item[4)] The orbital motion of the quartet is transformed into that of the center of mass signified by $N_\alpha L_\alpha$ and that of the relative motion between the two di-nucleons signified by $n_\alpha l_\alpha$.
\end{itemize}
The former three correspond to the $jj$ to $ls$ transformation, and the last corresponds to the Talmi-Moshinsky transformation. 
According to these transformation, the wave function of a quartet configuration writes as 
\begin{equation}
\begin{aligned}
\Phi_{J_\alpha M_\alpha}^{k_1k_2(j_{12})k_3k_4(j_{34})}
=& [ [ [\psi_{n_1l_1}\otimes \chi_{s_1}]_{j_1} \otimes [\psi_{n_2l_2}\otimes \chi_{s_2} ]_{j_2} ]_{j_{12}} \otimes [ [\psi_{n_3l_3}\otimes \chi_{s_3} ]_{j_3}\otimes [\psi_{n_4l_4}\otimes \chi_{s_4} ]_{j_4} ]_{j_{34}}]^{J_\alpha}_{M_\alpha}\\
=& \sum_{ l_{12} s_{12}  l_{34} s_{34}}\gamma^{(j_{12})}_{l_{12} s_{12}}(^{j_1 l_1 s_1}_{j_2 l_2 s_2}) \cdot \gamma^{(j_{34})}_{l_{34} s_{34}}(^{j_3 l_3 s_3}_{j_4 l_4 s_4}) \cdot \\
&[ [ [\psi_{n_1 l_1} \otimes \psi_{n_2 l_2}]_{n_{12} l_{12}} \otimes [\chi_{s_1} \otimes \chi_{s_2}]_{s_{12}}  ]_{j_{12}} \otimes [ [\psi_{n_3 l_3} \otimes \psi_{n_4 l_4}]_{n_{34} l_{34}} \otimes [\chi_{s_3} \otimes \chi_{s_4}]_{s_{34}}  ]_{j_{34}} ]^{J_\alpha}_{M_\alpha}\\
=& \sum_{ l_{12} s_{12}  l_{34} s_{34}} \sum_{l S} \gamma^{(j_{12})}_{l_{12} s_{12}}(^{j_1 l_1 s_1}_{j_2 l_2 s_2}) \cdot \gamma^{(j_{34})}_{l_{34} s_{34}}(^{j_3 l_3 s_3}_{j_4 l_4 s_4}) \cdot  \gamma^{(J_\alpha)}_{l S}(^{j_{12} l_{12} s_{12}}_{j_{34} l_{34} s_{34}}) \cdot \\
&[ [ [\psi_{n_1 l_1} \otimes \psi_{n_2 l_2}]_{n_{12} l_{12}} \otimes [\psi_{n_3 l_3} \otimes \psi_{n_4 l_4}]_{n_{34} l_{34}}  ]_{n l} \otimes [ [\chi_{s_1} \otimes \chi_{s_2}]_{s_{12}} \otimes [\chi_{s_3} \otimes \chi_{s_4}]_{s_{34}}  ]_{S} ]^{J_\alpha}_{M_\alpha}\\
=& \sum_{ l_{12} s_{12}  l_{34} s_{34}} \sum_{N_{12}\Lambda_{12} \eta_{12} \lambda_{12}} \sum_{N_{34}\Lambda_{34} \eta_{34} \lambda_{34}}\widetilde{\gamma}^{(j_{12}t_{12})}_{l_{12} s_{12}\lambda_{12}}(^{j_1 l_1 s_1}_{j_2 l_2 s_2}) \cdot \widetilde{\gamma}^{(j_{34}t_{34})}_{l_{34} s_{34}\lambda_{12}}(^{j_3 l_3 s_3}_{j_4 l_4 s_4}) \cdot \\
&\sum_{l S}  \gamma^{(J_\alpha)}_{l S}(^{j_{12} l_{12} s_{12}}_{j_{34} l_{34} s_{34}}) \cdot 
M_{l_{12}}(^{N_{12} \Lambda_{12} \eta_{12}\lambda_{12}}_{n_1l_1n_2l_2}) \cdot  M_{l_{34}}(^{N_{34} \Lambda_{34} \eta_{34}\lambda_{34}}_{n_3l_3n_4l_4}) \cdot  \\
&[ [ [\psi_{N_{12}\Lambda_{12}} \otimes \psi_{\eta_{12} \lambda_{12}}]_{n_{12} l_{12}} \otimes [\psi_{N_{34}\Lambda_{34}} \otimes \psi_{\eta_{34}\lambda_{34}}]_{n_{34} l_{34}}  ]_{n l} \otimes [ [\chi_{s_1} \otimes \chi_{s_2}]_{s_{12}} \otimes [\chi_{s_3} \otimes \chi_{s_4}]_{s_{34}}  ]_{S} ]^{J_\alpha}_{M_\alpha}\\
=& \sum_{ l_{12} s_{12}  l_{34} s_{34}} \sum_{N_{12}\Lambda_{12} \eta_{12} \lambda_{12}} \sum_{N_{34}\Lambda_{34} \eta_{34} \lambda_{34}} 
\widetilde{\gamma}^{(j_{12} t_{12})}_{l_{12} s_{12}\lambda_{12}}(^{j_1 l_1 s_1}_{j_2 l_2 s_2}) \cdot \widetilde{\gamma}^{(j_{34} t_{34})}_{l_{34} s_{34}\lambda_{34}}(^{j_3 l_3 s_3}_{j_4 l_4 s_4}) \cdot  \sum_{l S}\gamma^{(J_\alpha)}_{l S}(^{j_{12} l_{12} s_{12}}_{j_{34} l_{34} s_{34}}) \cdot \\
&\sum_{\Lambda \lambda} \gamma_{\Lambda \lambda}^{(l)}(^{l_{12} \Lambda_{12} \lambda_{12}}_{l_{34} \Lambda_{34} \lambda_{34}})  \cdot 
M_{l_{12}}(^{N_{12} \Lambda_{12} \eta_{12}\lambda_{12}}_{n_1l_1n_2l_2}) \cdot  M_{l_{34}}(^{N_{34} \Lambda_{34} \eta_{34}\lambda_{34}}_{n_3l_3n_4l_4}) \cdot  
\sum_{N_\alpha L_\alpha n_\alpha l_\alpha}
M_\Lambda(^{N_\alpha L_\alpha n_\alpha l_\alpha}_{N_{12}\Lambda_{12} N_{34}\Lambda_{34}}) \cdot \\
&[ [ [\psi_{N_\alpha L_\alpha} \otimes \psi_{n_\alpha l_\alpha}]_{N \Lambda} \otimes [\psi_{\eta_{12} \lambda_{12}} \otimes \psi_{\eta_{34}\lambda_{34}}]_{\eta \lambda}  ]_{n l} \otimes [ [\chi_{s_1} \otimes \chi_{s_2}]_{s_{12}} \otimes [\chi_{s_3} \otimes \chi_{s_4}]_{s_{34}}  ]_{S} ]^{J_\alpha}_{M_\alpha}\\
\end{aligned}
\end{equation}
where $\psi$ denotes the single nucleon wave function, $\chi$ denote the spin wave function, $l_{12 (34)} = l_{1(3)} \otimes l_{2 (4)}$ is the total orbital angular momentum of the nucleon-pair, 
$s_{12 (34)} = s_{1 (3)} \otimes s_{2 (4)}$ is the total intrinsic spin of the nucleon-pair, 
$t_{12 (34)} = t_{1 (3)} \otimes t_{2 (4)}$ is the total isospin of the nucleon-pair, 
$S = s_{12}\otimes s_{34}$ is the total intrinsic spin of the quartet, 
$l = l_{12}\otimes l_{34}$ is the total orbital angular momentum of the quartet, 
$\Lambda_{12 (34)}$ is the orbital angular momentum of the center of mass of the nucleon-pair,
$\lambda_{12 (34)}$ is the orbital angular momentum of the relative motion of nucleons in nucleon-pair,
$\Lambda = \Lambda_{12} \otimes \Lambda_{34}$ is the total orbital angular momentum of the two nucleon-pairs,
$\lambda = \lambda_{12} \otimes \lambda_{34}$ is the total orbital angular momentum of the two relative motions of nucleons in nucleon-pair,
$2N_{12 (34)} + \Lambda_{12 (34)} + 2\eta_{12 (34)} + \lambda_{12 (34)} = 2n_{1 (3)} + l_{1 (3)} + 2n_{2 (4)} + l_{2 (4)}$, 
$2N_\alpha + L_\alpha + 2n_\alpha + l_\alpha = 2N_{12} + \Lambda_{12} + 2N_{34} + \Lambda_{34}$,
$\gamma$ represents the $jj$ to $ls$ transformation coefficient and $\widetilde{\gamma}$ denotes the exchange asymmetric $jj$ to $ls$ transformation coefficient,
\begin{equation}
\gamma^{(j_{12})}_{l_{12} s_{12}}(^{j_1 l_1 s_1}_{j_2 l_2 s_2})=\sqrt{(2j_1+1)(2j_2+1)(2s_{12}+1)(2l_{12}+1)} 
\begin{Bmatrix}
l_1 & s_1 & j_1 \\
l_2 & s_2 & j_2 \\
l_{12} & s_{12} & j_{12}
\end{Bmatrix},
\end{equation}
\begin{equation}
\widetilde\gamma^{(j_{12}t_{12})}_{l_{12} s_{12}\lambda_{12}}(^{j_1 l_1 s_1}_{j_2 l_2 s_2}) = \frac{1- (-1)^{t_{12}+s_{12}+\lambda_{12}}}{\sqrt{2(1+\delta_{k_1k_2})}} \gamma^{(j_{12})}_{l_{12} s_{12}}(^{j_1 l_1 s_1}_{j_2 l_2 s_2}),
\end{equation}
$M$ is the Brody-Moshinsky Bracket \cite{brody_tables_1967, buck_simple_1996} which makes the Talmi-Moshinsky transformation, 
 \begin{equation}
\begin{aligned}
M_\lambda(^{n_3l_3n_4l_4}_{n_1l_1n_2l_2}) 
=&\langle n_3l_3,n_4l_4:\lambda|n_1l_1,n_2l_2:\lambda\rangle\\
=&\prod_{k=1}^{4}i^{-l_k}\sqrt{\frac{(n_k)![2(n_k+l_k)+1]!!}{\sqrt{2^{l_k}}}}\sum_{n_an_bn_cn_d,l_al_bl_cl_d}
 (-1)^{l_d} (\frac{1}{2})^{n_a+n_b+n_c+n_d}
\left\{\begin{matrix}
l_a & l_b & l_1\\
l_c & l_d & l_2\\
l_3 & l_4 & \lambda
\end{matrix}\right\}\\
&\langle l_a0l_c0|l_30\rangle \langle l_b0l_d0|l_40\rangle \langle l_a0l_b0|l_10\rangle \langle l_c0l_d0|l_20\rangle
\prod_{p=a}^d \frac{(-1)^{l_p}(2l_p+1)}{n_p![2(n_p+l_p)+1]!!}.
\end{aligned}
\end{equation} 
$\Phi_{J_\alpha M_\alpha}^{k_1k_2(j_{12})k_3k_4(j_{34})}$ is the wave function of the transferred four nucleons in the mother nucleus frame. In frame of $^{4}$He, the relative motion of each two nucleons is in $0s$ orbital, i.e., $\eta_{12}=\lambda_{12}=\eta_{34}=\lambda_{34}=0$. So the components becoming $\alpha$ particle in the quartet wave function is 
\begin{equation}
\begin{aligned}
\varPhi_{J_\alpha M_\alpha}^{k_1k_2(j_{12})k_3k_4(j_{34})}
=& \sum_{ l_{12} s_{12}  l_{34} s_{34}} \sum_{N_{12}\Lambda_{12} } \sum_{N_{34}\Lambda_{34} }  \widetilde\gamma^{(j_{12}t_{12})}_{l_{12} s_{12}}(^{j_1 l_1 s_1}_{j_2 l_2 s_2}) \cdot \widetilde\gamma^{(j_{34}t_{34})}_{l_{34} s_{34} }(^{j_3 l_3 s_3}_{j_4 l_4 s_4}) \cdot \sum_{l S} \gamma^{(J_\alpha)}_{l S}(^{j_{12} l_{12} s_{12}}_{j_{34} l_{34} s_{34}}) \cdot \\
&\sum_{\Lambda \lambda} \gamma_{\Lambda \lambda}^{(l)}(^{l_{12} \Lambda_{12} 0}_{l_{34} \Lambda_{34} 0})  \cdot M_{l_{12}}(^{N_{12} \Lambda_{12} 00}_{n_1l_1n_2l_2}) \cdot  M_{l_{34}}(^{N_{34} \Lambda_{34} 00}_{n_3l_3n_4l_4}) \cdot  
\sum_{N_\alpha L_\alpha n_\alpha l_\alpha}
M_\Lambda(^{N_\alpha L_\alpha n_\alpha l_\alpha}_{N_{12}\Lambda_{12} N_{34}\Lambda_{34}}) \cdot \\
=& \sum_{ l_{12} s_{12}  l_{34} s_{34}} \sum_{N_{12}\Lambda_{12} } \sum_{N_{34}\Lambda_{34} }  \widetilde\gamma^{(j_{12}t_{12})}_{l_{12} s_{12}}(^{j_1 l_1 s_1}_{j_2 l_2 s_2}) \cdot \widetilde\gamma^{(j_{34}t_{34})}_{l_{34} s_{34} }(^{j_3 l_3 s_3}_{j_4 l_4 s_4}) \cdot \sum_{l S} \gamma^{(J_\alpha)}_{l S}(^{j_{12} l_{12} s_{12}}_{j_{34} l_{34} s_{34}}) \cdot \\
&\sum_{\Lambda 0} \gamma_{\Lambda 0}^{(l)}(^{l_{12} \Lambda_{12} 0}_{l_{34} \Lambda_{34} 0})  \cdot M_{l_{12}}(^{N_{12} \Lambda_{12} 00}_{n_1l_1n_2l_2}) \cdot  M_{l_{34}}(^{N_{34} \Lambda_{34} 00}_{n_3l_3n_4l_4}) \cdot  
\sum_{N_\alpha L_\alpha n_\alpha l_\alpha}
M_\Lambda(^{N_\alpha L_\alpha n_\alpha l_\alpha}_{N_{12}\Lambda_{12} N_{34}\Lambda_{34}}) \cdot \\
&[ [ [\psi_{N_\alpha L_\alpha} \otimes \psi_{n_\alpha l_\alpha}]_{N \Lambda} \otimes [\psi_{0s} \otimes \psi_{0s}]_{0s}  ]_{nl} \otimes [ [\chi_{s_1} \otimes \chi_{s_2}]_{s_{12}} \otimes [\chi_{s_3} \otimes \chi_{s_4}]_{s_{34}}  ]_{S} ]^{J_\alpha}_{M_\alpha}\\
=& \sum_{ l_{12} s_{12}  l_{34} s_{34}} \sum_{N_{12}\Lambda_{12} } \sum_{N_{34}\Lambda_{34} }  \widetilde\gamma^{(j_{12}t_{12})}_{l_{12} s_{12}}(^{j_1 l_1 s_1}_{j_2 l_2 s_2}) \cdot \widetilde\gamma^{(j_{34}t_{34})}_{l_{34} s_{34} }(^{j_3 l_3 s_3}_{j_4 l_4 s_4}) \cdot \sum_{l S} \gamma^{(J_\alpha)}_{l S}(^{j_{12} l_{12} s_{12}}_{j_{34} l_{34} s_{34}}) \cdot \\
& \gamma_{l 0}^{(l)}(^{l_{12} \Lambda_{12} 0}_{l_{34} \Lambda_{34} 0})  \cdot M_{l_{12}}(^{N_{12} \Lambda_{12} 00}_{n_1l_1n_2l_2}) \cdot  M_{l_{34}}(^{N_{34} \Lambda_{34} 00}_{n_3l_3n_4l_4}) \cdot  
\sum_{N_\alpha L_\alpha n_\alpha l_\alpha}
M_l(^{N_\alpha L_\alpha n_\alpha l_\alpha}_{N_{12}\Lambda_{12} N_{34}\Lambda_{34}}) \cdot \\
&[ [ [\psi_{N_\alpha L_\alpha} \otimes \psi_{n_\alpha l_\alpha}]_{nl} \otimes [\psi_{0s} \otimes \psi_{0s}]_{0s}  ]_{nl} \otimes [ [\chi_{s_1} \otimes \chi_{s_2}]_{s_{12}} \otimes [\chi_{s_3} \otimes \chi_{s_4}]_{s_{34}}  ]_{S} ]^{J_\alpha}_{M_\alpha}\\
=& \sum_{ l_{12} s_{12}  l_{34} s_{34}}   \widetilde\gamma^{(j_{12}t_{12})}_{l_{12} s_{12}}(^{j_1 l_1 s_1}_{j_2 l_2 s_2}) \cdot \widetilde\gamma^{(j_{34}t_{34})}_{l_{34} s_{34} }(^{j_3 l_3 s_3}_{j_4 l_4 s_4}) \cdot \sum_{l S} \gamma^{(J_\alpha)}_{l S}(^{j_{12} l_{12} s_{12}}_{j_{34} l_{34} s_{34}}) \cdot 
\sum_{N_\alpha L_\alpha n_\alpha l_\alpha}M_l(^{N_\alpha L_\alpha n_\alpha l_\alpha}_{n_{12}l_{12} n_{34}l_{34}}) \cdot \\
&[ [ [\psi_{N_\alpha L_\alpha} \otimes \psi_{n_\alpha l_\alpha}]_{nl} \otimes [\psi_{0s} \otimes \psi_{0s}]_{0s}  ]_{nl} \otimes [ [\chi_{s_1} \otimes \chi_{s_2}]_{s_{12}} \otimes [\chi_{s_3} \otimes \chi_{s_4}]_{s_{34}}  ]_{S} ]^{J_\alpha}_{M_\alpha}\\
\end{aligned}
\end{equation}
The wave function of the transferred $\alpha$ particle in the mother nucleus frame includes the configuration mixing and thus express as  
\begin{equation}
\begin{aligned}
\varPhi_{J_\alpha M_\alpha} &= \sum_{k_1k_2k_3k_4 j_{12} j_{34}} \mathcal{S}_\alpha({J\{|^{k_1k_2(j_{12})}_{k_3k_4(j_{34})}(J_\alpha);J'})
\cdot \varPhi_{J_\alpha M_\alpha}^{k_1k_2(j_{12})k_3k_4(j_{34})}\\
&=\sum_{k_1k_2k_3k_4 j_{12} j_{34}} \mathcal{S}_\alpha({J\{|^{k_1k_2(j_{12})}_{k_3k_4(j_{34})}(J_\alpha);J'})
\cdot 
 \sum_{ l_{12} s_{12}  l_{34} s_{34}}   \widetilde\gamma^{(j_{12}t_{12})}_{l_{12} s_{12}}(^{j_1 l_1 s_1}_{j_2 l_2 s_2}) \cdot \widetilde\gamma^{(j_{34}t_{34})}_{l_{34} s_{34} }(^{j_3 l_3 s_3}_{j_4 l_4 s_4}) \cdot \sum_{l S} \gamma^{(J_\alpha)}_{l S}(^{j_{12} l_{12} s_{12}}_{j_{34} l_{34} s_{34}}) \cdot \\
&\sum_{N_\alpha L_\alpha n_\alpha l_\alpha}M_l(^{N_\alpha L_\alpha n_\alpha l_\alpha}_{n_{12}l_{12} n_{34}l_{34}}) \cdot 
[ [ [\psi_{N_\alpha L_\alpha} \otimes \psi_{n_\alpha l_\alpha}]_{nl} \otimes [\psi_{0s} \otimes \psi_{0s}]_{0s}  ]_{nl} \otimes [ [\chi_{s_1} \otimes \chi_{s_2}]_{s_{12}} \otimes [\chi_{s_3} \otimes \chi_{s_4}]_{s_{34}}  ]_{S} ]^{J_\alpha}_{M_\alpha}\\
\end{aligned}
\end{equation}
The probability density distribution of the center-of-mass of the $\alpha$ particle is 
\begin{equation}
\label{eq:P_formation_th}
\begin{aligned}
P(R_\alpha) =& \int d\chi_{s_1} d\chi_{s_2} d\chi_{s_3} d\chi_{s_4} dV_{r_{12}} dV_{r_{34}} dV_{r} dS_{R_\alpha} \varPhi_{J_\alpha M_\alpha}^2\\
  =&\sum_{k_1k_2k_3k_4 j_{12} j_{34}} \mathcal{S}_\alpha({J\{|^{k_1k_2(j_{12})}_{k_3k_4(j_{34})}(J_\alpha);J'}) \cdot 
\sum_{ l_{12} s_{12}  l_{34} s_{34}}   \widetilde\gamma^{(j_{12}t_{12})}_{l_{12} s_{12}}(^{j_1 l_1 s_1}_{j_2 l_2 s_2}) \cdot \widetilde\gamma^{(j_{34}t_{34})}_{l_{34} s_{34} }(^{j_3 l_3 s_3}_{j_4 l_4 s_4}) \cdot \sum_{l S} \gamma^{(J_\alpha)}_{l S}(^{j_{12} l_{12} s_{12}}_{j_{34} l_{34} s_{34}}) \cdot \\
&\sum_{k_1'k_2'k_3'k_4' j_{12}' j_{34}'} \mathcal{S}_\alpha({J\{|^{k_1'k_2'(j_{12}')}_{k_3'k_4'(j_{34}')}(J_\alpha);J'}) \cdot 
\sum_{ l_{12}'   l_{34}' }   \widetilde\gamma^{(j_{12}'t_{12}')}_{l_{12}' s_{12}}(^{j_1' l_1' s_1}_{j_2' l_2' s_2}) \cdot \widetilde\gamma^{(j_{34}'t_{34}')}_{l_{34}' s_{34} }(^{j_3' l_3' s_3}_{j_4' l_4' s_4}) \cdot \sum_{l' } \gamma^{(J_\alpha)}_{l' S}(^{j_{12}' l_{12}' s_{12}}_{j_{34}' l_{34}' s_{34}}) \cdot \\
& \sum_{N_\alpha L_\alpha n_\alpha l_\alpha}M_l(^{N_\alpha L_\alpha n_\alpha l_\alpha}_{n_{12}l_{12} n_{34}l_{34}})  \cdot 
 M_{l'}(^{N_\alpha L_\alpha n_\alpha l_\alpha}_{n_{12}'l_{12}' n_{34}'l_{34}'}) \cdot 
  \psi_{N_\alpha L_\alpha}^2  R_\alpha^2\\
  =&\sum_{k_{i,i\in[1,4]}, j_{12}, j_{34}} \sum_{k_{i,i\in[1,4]}', j_{12}', j_{34}'} \mathcal{S}_\alpha({J\{|^{k_1k_2(j_{12})}_{k_3k_4(j_{34})}(J_\alpha);J'}) \cdot 
  \mathcal{S}_\alpha({J\{|^{k_1'k_2'(j_{12}')}_{k_3'k_4'(j_{34}')}(J_\alpha);J'}) \cdot \\
&\sum_{N_\alpha n_\alpha l_\alpha} \sum_{l_{12}s_{12}l_{34}s_{34}} \sum_{l_{12}'s_{12}'l_{34}'s_{34}'} \mathcal{M}_\alpha(^{N_\alpha L_\alpha} _{n_\alpha l_\alpha} \{|^{k_1k_2(j_{12}l_{12}s_{12})}_{k_3k_4(j_{34}l_{34}s_{34})} \cdot  \mathcal{M}_\alpha(^{N_\alpha L_\alpha} _{n_\alpha l_\alpha} \{|^{k_1'k_2'(j_{12}'l_{12}'s_{12}')}_{k_3'k_4'(j_{34}'l_{34}'s_{34}')} \cdot 
\psi_{N_\alpha L_\alpha}^2(R_\alpha)
\end{aligned}
\end{equation}
where the center-of-mass amplitude $\mathcal{M}_\alpha$ writes as
\begin{equation}
\mathcal{M}_\alpha(^{N_\alpha L_\alpha} _{n_\alpha l_\alpha} \{|^{k_1k_2(j_{12}l_{12}s_{12})}_{k_3k_4(j_{34}l_{34}s_{34})}) 
=  \widetilde\gamma^{(j_{12}t_{12})}_{l_{12} s_{12}}(^{j_1 l_1 s_1}_{j_2 l_2 s_2}) \cdot \widetilde\gamma^{(j_{34}t_{34})}_{l_{34} s_{34} }(^{j_3 l_3 s_3}_{j_4 l_4 s_4}) \cdot \gamma^{(J_\alpha)}_{J_\alpha 0}(^{j_{12} l_{12} s_{12}}_{j_{34} l_{34} s_{34}}) \cdot 
M_l(^{N_\alpha L_\alpha n_\alpha l_\alpha}_{n_{12}l_{12} n_{34}l_{34}}).
\end{equation}

When matching with the decay process, the center-of-mass motion of the $\alpha$ particle has to be transformed to the relative motion between the $\alpha$ particle and the daughter nucleus. 
Noting $\vec{R}_\alpha$ and $\vec{R}_d$ the radial coordinates of the $\alpha$ particle and the daughter nucleus in the frame of the mother nucleus,
the coordinate of their relative motion is $\vec{R} = \vec{R}_\alpha - \vec{R}_d$.
To match the theoretical value of $P(R_\alpha)$ with the experimental value of $|R\Psi_{L_\alpha}|^2$, one needs to make a connection between $\vec{R}_\alpha$ and $\vec{R}$.
In frame of the mother nucleus, the ceter-of-mass of the $\alpha$ particle and the daughter nucleus is at rest, thus
\begin{equation}
\frac{m_\alpha \vec{R}_\alpha + m_d \vec{R}_d}{m_\alpha + m_d} = 0,
\end{equation}
providing 
\begin{equation}
  \vec{R}_d = - \frac{m_\alpha}{m_d} \vec{R}_\alpha \approx -\frac{A_\alpha}{A_d} \vec{R}_\alpha 
\end{equation}
where $A_\alpha$ and $A_d$ are the mass number of the $\alpha$ particle and the daughter nucleus,
and
\begin{equation}
  \vec{R} = \frac{A_\alpha + A_d}{A_d} \vec{R}_\alpha,
\end{equation} 
which means 
\begin{equation}
  |R\Psi_{L_\alpha}(R)|^2 = P(R_\alpha)
\end{equation}

Another thing should be counted is the form of the wavefunction of the ceter-of-mass motion of the $\alpha$ particle, i.e., the form of $\psi_{N_\alpha L_\alpha}$ in $P(R_\alpha)$.
The configuration-interaction shell model takes harmonic oscillator wave function as the single nucleon basis.
The radial part of the harmonic oscillator wave function is 
\begin{equation}
  \psi_{nl}(r) = \sqrt{\frac{2^{l-n+2}(2(n+l)+1)!!\alpha^{2l+3}}{\sqrt{\pi}n![(2l+1)!!]^2}}\times \exp(-\frac{1}{2}\alpha^2r^2) \times r^l \times \sum_{k=0}^n \frac{(-1)^k 2^k n! (2l+1)!!(\alpha^2 r^2)^k}{k!(n-k)!(2(l+k)+1)!!},
\end{equation}
where $\alpha^2=m\omega/\hbar$ is the form factor.
The Hamiltonian describing the independent motion of four particles in a harmonic oscillator potential is  
\begin{equation}
\begin{aligned}
H &= \frac{p_1^2}{2m} + \frac{1}{2}m\omega^2r_1^2 + \frac{p_2^2}{2m} + \frac{1}{2}m\omega^2r_2^2 +  \frac{p_3^2}{2m} + \frac{1}{2}m\omega^2r_3^2 +\frac{p_4^2}{2m} + \frac{1}{2}m\omega^2r_4^2 \\
&= \frac{P_{12}^2+p_{12}^2}{2m} + \frac{P_{34}^2+p_{34}^2}{2m} + \frac{1}{2}m\omega^2(R_{12}^2 + r_{12}^2)  + \frac{1}{2}m\omega^2(R_{34}^2+r_{34}^2)\\
&= \frac{P_{12}^2+P_{34}^2}{2m} + \frac{p_{12}^2+p_{34}^2}{2m} + \frac{1}{2}m\omega^2(R_{12}^2 + R_{34}^2)  + \frac{1}{2}m\omega^2(r_{12}^2+r_{34}^2)\\
&= \frac{P_{1234}^2+p_{12-34}^2}{2m} + \frac{p_{12}^2+p_{34}^2}{2m} + \frac{1}{2}m\omega^2(R_{1234}^2 + r_{12-34}^2)  + \frac{1}{2}m\omega^2(r_{12}^2+r_{34}^2)\\
\end{aligned}
\end{equation}
where $\vec{P}_{12} = \frac{\vec{p}_1+\vec{p}_2}{\sqrt2}$, 
$\vec{P}_{34} = \frac{\vec{p}_3+\vec{p}_4}{\sqrt2}$, 
$\vec{P}_{1234} = \frac{\vec{P}_{12}+\vec{P}_{34}}{\sqrt2} = \frac{\vec{p}_1 + \vec{p}_2 + \vec{p}_3 + \vec{p}_4}{2}$, 
$\vec{p}_{12} = \frac{\vec{p}_1-\vec{p}_2}{\sqrt2}$, 
$\vec{p}_{34} = \frac{\vec{p}_3-\vec{p}_4}{\sqrt2}$, 
$\vec{p}_{12-34} = \frac{\vec{p}_{12}+\vec{p}_{34}}{\sqrt2}$, 
$\vec{R}_{12} = \frac{\vec{r}_1 + \vec{r}_2}{\sqrt2}$,
$\vec{R}_{34} = \frac{\vec{r}_3 + \vec{r}_4}{\sqrt2}$,
$\vec{R}_{1234} = \frac{\vec{R}_{12}+\vec{R}_{34}}{\sqrt2} = \frac{\vec{r}_1 + \vec{r}_2 + \vec{r}_3 + \vec{r}_4}{2}$,
$\vec{r}_{12} = \frac{\vec{r}_1 - \vec{r}_2}{\sqrt2}$,
$\vec{r}_{34} = \frac{\vec{r}_3 - \vec{r}_4}{\sqrt2}$,
$\vec{r}_{12-34} = \frac{\vec{r}_{12}-\vec{r}_{34}}{\sqrt2}$. 
Taking out the part of the center-of-mass motion, the Hamiltonian is
\begin{equation}
  \begin{aligned}
  H &= \frac{P_{1234}^2}{2m} + \frac{1}{2}m\omega^2 R_{1234}^2\\
    &= \frac{1}{2m}(\frac{\vec{p}_1 + \vec{p}_2 + \vec{p}_3 + \vec{p}_4}{2})^2 + \frac{1}{2}m\omega^2 (\frac{\vec{r}_1 + \vec{r}_2 + \vec{r}_3 + \vec{r}_4}{2})^2\\
    &= \frac{1}{2(4m)}(\vec{p}_1 + \vec{p}_2 + \vec{p}_3 + \vec{p}_4)^2 + \frac{1}{2}(4m)\omega^2 (\frac{\vec{r}_1 + \vec{r}_2 + \vec{r}_3 + \vec{r}_4}{4})^2\\
  \end{aligned}
\end{equation}
So the form factor of the center-of-mass motion of four-nucleon system is four times that of the single particle.
The radial part of the wave function of the center-of-mass motion of the four-nucleon system is thus
\begin{equation}
  \psi_{nl}(r) = \sqrt{\frac{2^{l-n+2}(2(n+l)+1)!!\alpha'^{2l+3}}{\sqrt{\pi}n![(2l+1)!!]^2}}\times \exp(-\frac{1}{2}\alpha'^2r^2) \times r^l \times \sum_{k=0}^n \frac{(-1)^k 2^k n! (2l+1)!!(\alpha'^2 r^2)^k}{k!(n-k)!(2(l+k)+1)!!},
\end{equation}
where $\alpha'^2=4m\omega/\hbar$. As to $\omega$, the Blomqvist–Molinari formula $\hbar\omega=45 A^{-1/3} - 25 A^{-2/3}$ MeV is generally used \cite{blomqvist_collective_1968}.

\subsection{III. Decay process: Surface Emergence and Penetration}\label{sec:mt:sfp}

Assuming that the $\alpha$ particle is formed inside the nucleus with  the orbital angular momentum $L_\alpha$ and wave $\Psi_{L_\alpha}$, the decay width for the emission of the preformed particle from the surface could be expressed as
\begin{equation}
\Gamma_{L_\alpha} = \frac{\hbar}{T_\alpha} = \frac{\hbar^2R}{\mu} |\Psi_{L_\alpha}(R)|^2 \frac{kR}{|H^+_{L_\alpha}(kR)|^2},
\label{eq:thomas}
\end{equation} 
where $T_\alpha$ is the half-life, $\mu$ is the reduced mass of the two-body system consisting of the $\alpha$-particle and the daughter nucleus, $k=\sqrt{2\mu Q_\alpha}/\hbar$ is the wave number, $Q_\alpha$ is the decay energy, $R$ is the distance between the alpha particle on the surface and the center of mass of the daughter nucleus which is taken here as $R=r_0(A_\alpha^{1/3}+A_d^{1/3})$, $A_\alpha$ and $A_d$ are the mass number of the $\alpha$ particle and the daughter nucleus, $H^+_{L_\alpha}$ is the outgoing Coulomb-Hankel function. 

One may derive  Eq. (\ref{eq:thomas}) with the aid of probability flux density $\vec j$.
The continuity equation is
\begin{equation}
\frac{\partial (\psi^*\psi)}{\partial t} = \frac{i\hbar}{2\mu}(\psi^*\triangle\psi-\psi\triangle\psi^*) = \frac{i\hbar}{2\mu}\vec\triangledown\cdot(\psi^*\vec\triangledown\psi - \psi\vec\triangledown\psi^*) = -\vec\triangledown \cdot \vec{j}.
\end{equation}
The variation rate of probability that the $\alpha$ particle exists in a volume $V$ equals to the probability flux passing through the closed surface $S$ of the volume $V$, which is also the frequency that the $\alpha$ particle passing through the closed surface $S$:
\begin{equation}
f_\alpha= \frac{\partial \int \psi^*\psi dV}{\partial t}   = -\int \vec\triangledown\cdot \vec{j} dV = -\oint \vec{j} \cdot d\vec{S}.
\end{equation}
When the $\alpha$ particle is far outside the residue nucleus, its motion is just governed as an outgoing Coulomb wave
\begin{equation}
\psi^+_{L_\alpha}(r, \theta, \varphi) = \frac{\mathcal{N}_{L_\alpha}H^+_{L_\alpha}(kr)Y^{L_\alpha}_{M}(\theta,\varphi)}{r},
\end{equation}
where $\mathcal{N}_{L_\alpha}$ is the matching factor and $Y^{L_\alpha}_M(\theta,\varphi)$ is the spherical harmonic function. The probability flux density then is 
\begin{equation}
\begin{aligned}
\vec{j}=& -\frac{i\hbar}{2\mu}(\psi^{+*}_{L_\alpha}\vec\triangledown\psi^+_{L_\alpha} - \psi^+_{L_\alpha}\vec\triangledown\psi^{+*}_{L_\alpha})\\
=&-\frac{i\hbar}{2\mu}\mathcal{N}_{L_\alpha}^2\bigg[ \frac{1}{r^2}\big[H^-_{L_\alpha}(kr)\frac{\partial H^+_{L_\alpha}(kr)}{\partial r}-H^+_{L_\alpha}(kr)\frac{\partial H^-_{L_\alpha}(kr)}{\partial r}\big]Y^{L_\alpha}_{M}(\theta,\varphi)Y^{L_\alpha}_{-M}(\theta,\varphi)\vec{e}_r+\\
&\frac{H^+_{L_\alpha}(kr)H^-_{L_\alpha}(kr)}{r^3}\big[Y^{L_\alpha}_{-M}(\theta,\varphi)\frac{\partial Y^{L_\alpha}_M(\theta,\varphi)}{\partial \theta} - Y^{L_\alpha}_{M}(\theta,\varphi)\frac{\partial Y^{L_\alpha}_{-M}(\theta,\varphi)}{\partial \theta}\big] \vec{e}_\theta + \\
&\frac{H^+_{L_\alpha}(kr)H^-_{L_\alpha}(kr)}{r^3\sin\theta}iM\big[Y^{L_\alpha}_{-M}(\theta,\varphi)Y^{L_\alpha}_M(\theta,\varphi) + Y^{L_\alpha}_{M}(\theta,\varphi)Y^{L_\alpha}_{-M}(\theta,\varphi)\big]\vec{e}_\varphi\bigg].
\end{aligned}
\end{equation}
We can concentrate only on the radial part
\begin{equation}
\begin{aligned}
j_r &=-\frac{i\hbar}{2\mu}\mathcal{N}_{L_\alpha}^2 \frac{Y^{L_\alpha}_{M}(\theta,\varphi)Y^{L_\alpha}_{-M}(\theta,\varphi)}{r^2}\big[H^-_{L_\alpha}(kr)\frac{\partial H^+_{L_\alpha}(kr)}{\partial r}-H^+_{L_\alpha}(kr)\frac{\partial H^-_{L_\alpha}(kr)}{\partial r}\big]\\
&= \mathcal{N}_{L_\alpha}^2 \frac{\hbar k}{\mu}\frac{Y^{L_\alpha}_{M}(\theta,\varphi)Y^{L_\alpha}_{-M}(\theta,\varphi)}{r^2}
\big[G_{L_\alpha}(kr)F_{L_\alpha}'(kr) - F_{L_\alpha}(kr)G_{L_\alpha}'(kr) \big]\\
\end{aligned}
\end{equation}
Considering the asymptotic behavior of $F_{L_\alpha}$ and $G_{L_\alpha}$ at large $\rho=kr$ \cite{olver2010}:
\begin{equation}
\begin{aligned}
F_{L_\alpha} &= g\cos \theta_{L_\alpha} + f\sin\theta_{L_\alpha}\\
G_{L_\alpha} &= f\cos \theta_{L_\alpha} - g\sin\theta_{L_\alpha}\\
F_{L_\alpha}' &= \hat g \cos \theta_{L_\alpha} + \hat f\sin\theta_{L_\alpha}\\
G_{L_\alpha}' &= \hat f\cos \theta_{L_\alpha} - \hat g\sin\theta_{L_\alpha}\\
\end{aligned}
\end{equation}
with 
\begin{equation}
g\hat f - f \hat g = 1,
\end{equation}
one can get
\begin{equation}
\begin{aligned}
G_{L_\alpha}F_{L_\alpha}' - F_{L_\alpha}G_{L_\alpha}' = &(f\cos \theta_{L_\alpha} - g\sin\theta_{L_\alpha})(\hat g \cos \theta_{L_\alpha} + \hat f\sin\theta_{L_\alpha})
-(g\cos \theta_{L_\alpha} + f\sin\theta_{L_\alpha})(\hat f\cos \theta_{L_\alpha} - \hat g\sin\theta_{L_\alpha})\\
= &(f\hat g - g \hat f) \cos^2\theta_{L_\alpha} +(f \hat g - g\hat f)\sin^2\theta_{L_\alpha} + [(f\hat f - g \hat g) 
-(f\hat f - g \hat g) ]\sin\theta_{L_\alpha}\cos\theta_{L_\alpha}\\
 =& f\hat g - g \hat f \\
=&-1,
\end{aligned}
\end{equation}
and
\begin{equation}
j_{r}(kr\to\infty) = -\mathcal{N}_{L_\alpha}^2 \frac{\hbar k}{\mu}\frac{Y^{L_\alpha}_{M}(\theta,\varphi)Y^{L_\alpha}_{-M}(\theta,\varphi)}{r^2}
\end{equation}
Finally, the integration over a closed spherical surface gives
\begin{equation}
\begin{aligned}
f = -\int \vec{j}\cdot \vec{dS}  &= \mathcal{N}_{L_\alpha}^2 \frac{\hbar k}{\mu}\int dS \frac{Y^{L_\alpha}_{M}(\theta,\varphi)Y^{L_\alpha}_{-M}(\theta,\varphi)}{r^2}\\
&= \mathcal{N}_{L_\alpha}^2 \frac{\hbar k }{\mu} \int d\theta d\varphi r^2 \sin\theta \sin\varphi \frac{Y^{L_\alpha}_{M}(\theta,\varphi)Y^{L_\alpha}_{-M}(\theta,\varphi)}{r^2}\\
&= \mathcal{N}_{L_\alpha}^2 \frac{\hbar k }{\mu}
\end{aligned}
\end{equation}

To determine the matching factor $\mathcal{N}_{L_\alpha}$, the Coulomb wave function $\psi^+_{L_\alpha}$ has to be matched with the initial wave $\Psi_{L_\alpha}$ at $r=R$:
\begin{equation}
\frac{\mathcal{N}_{L_\alpha}H^+_{L_\alpha}(kR)}{R} = \Psi_{L_\alpha}(R),
\end{equation}
which gives
\begin{equation}
\mathcal{N}_{L_\alpha} = \frac{R\Psi_{L_\alpha}(R)}{H^+_{L_\alpha}(kR)},
\end{equation}
and thus
\begin{equation}
\Gamma_{L_\alpha} = \hbar /T_{\alpha} = \hbar f = \frac{\hbar^2 R}{\mu} |\Psi_{L_\alpha}(R)|^2 \frac{kR}{|H^+_{L_\alpha}(kR)|^2}. 
\end{equation}
In this way, the formation probability can be extracted experimentally through:
\begin{equation}
\label{f_exp}
|\mathcal{F}_{L_\alpha}|^2_\mathrm{expt.} = |\Psi_{L_\alpha}(R)|^2 = \frac{\mu }{\hbar^2 R} \frac{|H_{L_\alpha}^+(kR)|^2}{kR} \Gamma_{L_\alpha}.
\end{equation}

\end{widetext}

\end{document}